\documentclass[12pt]{article}

\usepackage{graphicx}
\usepackage{qcmc06}
\usepackage{fancyhdr}

\newcommand{\beq}{\vspace{-1.5mm}\begin{equation}}
\newcommand{\eeq}{\vspace{-1.5mm}\end{equation}}
\newcommand{\beqa}{\begin{eqnarray}}
\newcommand{\eeqan}{\end{eqnarray*}}
\newcommand{\beqan}{\begin{eqnarray*}}
\newcommand{\eeqa}{\end{eqnarray}}
\newcommand{\bra}[1]{\langle{#1}|}

\newcommand{\ket}[1]{|{#1}\rangle}

\def\openone{{\leavevmode\hbox{\small1\kern-4pt\normalsize1}}}
\newcommand{\id}{\openone}
\newcommand{\ip}[1]{\langle{#1}\rangle}

\newcommand{\eqr}[1]{Eq.~(\ref{#1})}

\newcommand{\tr}{{\rm Tr}}

\begin{document}

\pagestyle{fancy}
 \fancyhead{}
 \setcounter{page}{421}
 \fancyfoot[c]{--\ \thepage\ --}

\title{Group theoretic formulation of complementarity}

\author{J.A. Vaccaro}

\affiliation{Centre for Quantum Computer Technology, Centre for Quantum Dynamics,
School of Biomolecular and Physical Sciences, Griffith University, Brisbane 4111, Australia}

\begin{abstract}

We generalize Bohr's complementarity principle for wave and particle properties to arbitrary
quantum systems. We begin by noting that a particle-like state is represented by a
spatially-localized wave function and its narrow probability density is displaced by spatial
translations. In contrast a wave-like state is represented by a spatially-delocalized wave function
and the corresponding broad position probability density is invariant to spatial translations. The
wave-particle dichotomy can therefore be seen as a competition between displacement and invariance
of the state with respect to spatial translations. We generalize this dichotomy to arbitrary
quantum systems with finite dimensional Hilbert spaces as follows. We use arbitrary finite symmetry
groups to represent transformations of the quantum system. The symmetry (i.e. invariance) or
asymmetry (i.e. displacement) of a given state with respect to transformations of the group are
identified with the generalized wave and particle nature, respectively. We adopt a measure of wave
and particle properties based on the amount of information that can be encoded in the symmetric and
asymmetric parts of the state.

\end{abstract}

\setcounter{section}{0}

\section{Introduction}

Bohr's complementarity principle is a defining feature of quantum physics \cite{bohr}. In essence
it represents the dichotomy between the particle and wave nature of mechanical objects; the
particle properties are typically symbolized by well-defined position and the wave properties by
well-defined momentum. More recent work has attempted to quantify the wave-like and particle-like
properties and study the range of properties in between the extremes of pure wave- and pure
particle-like states.  For example Wootters and Zurek \cite{WooZur} formulated an inequality for a
double slit experiment that expresses a lower bound on the loss of path information (i.e.
information about which slit a photon passes through) for a given sharpness of the interference
pattern. Scully {\em et al.} \cite{ScuEng} explored the erasure of path information and the
recovery of an interference pattern using sub-ensembles conditioned on ancillary measurements. A
debate regarding the application of an uncertainty principle ensured (\cite{Wise} and references
therein).  Also Englert \cite{Eng} derived an inequality for a two-way interferometer that limits
the distinguishability of the outcomes of a path measurement and the visibility of the interference
pattern. The related study of the simultaneous measurement of non-commuting observables also has a
similarly long history \cite{Ozawa}.

Here we generalize the complementarity principle as follows.  We first note that  a state with
particle-like properties is {\em displaced} (i.e. mapped) under a spatial translation to an
orthogonal state, owing to its narrowness in the position representation. For example, the wave
function $\psi(x)=\ip{x|\psi}$, where $\ket{x}$ is an eigenket of position and $\ket{\psi}$ is a
general state, is mapped to $\bra{x}e^{i\hat p\delta x/\hbar}\ket{\psi}=\ip{x+\delta
x|\psi}=\psi(x+\delta x)$ under the spatial translation given by $e^{i\hat p\delta x/\hbar}$ where
$\hat p$ is the position operator. The overlap $\int \psi^*(x)\psi(x+\delta x) dx$ is negligible
for finite translations $\delta x>0$ and particle-like states of the kind $\ket{\psi}\propto\int
e^{-(x-x')^2/4\sigma^2}\ket{x}dx$ with a sufficiently small value of $\sigma$.  In contrast, a
wave-like state is essentially delocalized in the position representation so that the position
probability density $P_{\rm pos}(x)=|\ip{x|\psi}|^2$ is essentially ``flat'' and {\em invariant} to
spatial translations.  For example, $P_{\rm pos}(x+\delta x)=|\ip{x|e^{i\hat p\delta
x/\hbar}|\psi}|^2$ is approximately equal to $P_{\rm pos}(x)$ for arbitrary translations $\delta x$
and wave-like states of the kind $\ket{\psi}\propto\int e^{-(x-x')^2/4\sigma^2}\ket{x}dx$ with a
sufficiently large value of $\sigma$. In other words, waves are {\it symmetric} (i.e. invariant)
and particles are {\it asymmetric} (i.e. displaced) with respect to spatial translations.

Next we generalize this notion by associating ``generalized'' wave and particle nature with
symmetry and asymmetry with respect to an arbitrary finite symmetry group, which we represent as
$G=\{g_1, g_2, \ldots, g_{|G|}\}$ of order $|G|$. We consider the generalized particle and wave
nature of a system with state density operator $\hat\rho$. For convenience we call the generalized
particles simply ``particles'', and similarly generalized waves ``waves'', and we refer to
transformations by the group as ``translations''. Let $G$ have the unitary representation $\hat
T_g$ for $g\in G$ on the system's Hilbert space.
\section{Information theoretic complementarity}

We first consider particle nature of the state $\hat\rho$.  Particle-like states are translated by
the actions of the group $G=\{g\}$ and hence their translation
\beq
    \hat\rho\mapsto\hat\rho_g = \hat T_g\hat\rho \hat T_g^\dagger
\eeq
for $g\in G$ can carry information. We imagine an information theoretic scenario between two
separated parties A and B as follows. Party A prepares the system in the translated state
$\hat\rho_g$ for $g\in G$ with uniform probability $p(g)=\frac{1}{|G|}$ and sends it to B.  Party B
then makes a measurement on the system to estimate the value of the parameter $g$.  We define an
{\it information-theoretic measure of particle nature}, $N_{\rm Part} (\hat\rho)$, of $\hat\rho$ as
the maximum of the mutual information between A and B over all possible measurements at
B\footnote[4]{In other words, $N_{\rm Part}(\hat\rho)$ is the {\em accessible} information B has
about the parameter $g$ prepared by A.}. Let B make the measurement described by the Kraus
operators $\hat M_k$ with POM elements $\hat\pi_k=\hat M_k^\dagger\hat M_k$ satisfying
$\sum_k\hat\pi=\hat\id$. The probability that B obtains outcome $k$ given that the system is
initially in state $\hat\rho_g$ is $P(k|g)=\tr(\hat M_k\hat\rho_g\hat M_k^\dagger)$. Note that
$p(g) P(k|g)= q(k)Q(g|k)$, where $q(k)=\sum_g p(g)P(k|g)$ is the average probability of B obtaining
result $k$, and $Q(g|k)$ is the probability the system was prepared in state $\hat\rho_g$ given
that B obtains the outcome $k$. Thus we find mutual information shared by both A and B for this
measurement is given by
\beq
   I_{\rm Part} = H(\{p_g\}) - \sum_k q(k) H(\{Q(g|k)\})
\eeq
where the $H(\{r(j)\})$ is the Shannon entropy associated with the set of probabilities $\{r(j):
j=1,2,\ldots \}$, i.e.
\beq
    H(\{r(j)\})=-\sum_j r(j)\log r(j)\ .
\eeq
The particle nature $N_{\rm Part}(\hat\rho)$ of $\hat\rho$ is the maximum of $I_{\rm Part}$ over
all possible measurements at B. It is bounded above by Holevo's theorem \cite{Schumacher}:
\beq
  \label{N_P_bound}
   N_{\rm Part}(\hat\rho)\le S({\cal G}[\hat\rho])-\frac{1}{|G|}\sum_{g\in G} S(\hat\rho_g)
\eeq
where $S(\hat\varrho)=-{\rm Tr}(\hat\varrho\ln\hat\varrho)$ is the von Neumann entropy of
$\hat\varrho$, and \cite{BarWis}
\beq
   {\cal G}[\hat\rho]\equiv\frac{1}{|G|}\sum_{g\in G} \hat T_g \hat\rho \hat T_g^\dagger
   =\frac{1}{|G|}\sum_{g\in G} \hat\rho_g
\eeq
is the average state received by B. Noting that as $\hat T_g$ is unitary, $S(\hat\rho_g)=S(\hat
T_g\hat\rho\hat T_g^\dagger)=S(\hat\rho)$ for all $g\in G$ we find \eqr{N_P_bound} becomes
\beq
  \label{N_part}
   N_{\rm Part}(\hat\rho)\le S({\cal G}[\hat\rho])-S(\hat\rho)\ .
\eeq
We have previously defined the quantity $A_G(\hat\rho)=S({\cal G}[\hat\rho])-S(\hat\rho)$ as the
asymmetry of $\hat\rho$ with respect to the group $G$ \cite{Vacc}; hence we find here that {\it the
information-theoretic measure of particle nature is bounded by the asymmetry of $\hat\rho$}, which
is consistent with our identification of particles with asymmetry.

We now consider the analogous information theoretic scenario for wave nature. We imagine that party
A encodes information in the wave properties of the state $\hat\rho$, using a suitably restricted
class of operations that leave the particle properties unchanged, and then sends the system to B
who decodes the information using measurements. Rather than specify the kinds of operators that A
can use we note that the wave properties of the state are invariant to translations $\hat T_g$ for
$g\in G$, and so in terms of the wave nature the state $\hat\rho$ is equivalent to $\hat
T_g\hat\rho \hat T_g^\dagger$ and also to ${\cal G}[\hat\rho]$. Moreover, the latter state ${\cal
G}[\hat\rho]$ is symmetric in the sense that it is invariant to translations of the group
\cite{BarWis}, i.e.
\beq
   \hat T_g\Big({\cal G}[\hat\rho]\Big)\hat T_g^\dagger = {\cal G}[\hat\rho]\ \ {\rm for\ }g\in G\ ,
\eeq
and so ${\cal G}[\hat\rho]$ is devoid of any particle nature that $\hat\rho$ might have.  Thus A
can encode information using {\em arbitrary} operations on ${\cal G}[\hat\rho]$ and be sure that
the encoding uses only wave properties of $\hat\rho$.  This leads to an equivalent measure for wave
nature as follows. Party A encodes information in the system by preparing the state
\beq
   \hat\rho'_j=\hat U_j\Big({\cal G}[\hat\rho]\Big)\hat U_j^\dagger
\eeq
with probability $p'(j)=\frac{1}{N}$ for an arbitrary set of $N$ unitary operators $\{\hat U_1,
\hat U_2, \ldots, \hat U_N\}$.  The system is then sent to B who makes a measurement to estimate
the value of the parameter $j$.  Let B's measurement be described by the Kraus operators $\hat
M'_k$ with POM elements $\hat\pi'_k=\hat M_k^{\prime\dagger} \hat M'_k$ where
$\sum_k\hat\pi'_k=\hat\id$.  The probability that B obtains outcome $k$ for the system in state
$\hat\rho'_j$ is given by $P'(k|j)=\tr(\hat M'_k\hat\rho'_j\hat M_k^{\prime\dagger})$.  Again note
that  $p'(j)P'(k|j)= q'(k)Q'(j|k)$ where $q'(k)=\sum_j p'(j) P'(k|j)$ is the average probability of
B obtaining result $k$ and $Q'(j|k)$ is the probability the system was prepared in state
$\hat\rho'_j$ given that B obtains the outcome $k$. Thus the mutual information shared by both A
and B for this measurement is given by
\beq
   I_{\rm Wave} = H(\{p'(j)\}) - \sum_k q'(k) H(\{Q'(j|k)\})\ .
\eeq
The maximum of $I_{\rm Wave}$ over all possible measurements at B is bounded by Holevo's theorem
\cite{Schumacher}:
\beq
  \label{I_W_max}
   I_{\rm Wave}^{\rm (max)}\le S\Big(\sum_j p'(j)\hat\rho'_j\Big)-\sum_j p'(j)S(\hat\rho'_j)\ .
\eeq
As the operators $\hat U_j$ are unitary we find $S(\hat\rho'_j)=S({\cal G}[\hat\rho])$.  We define
an {\it information-theoretic measure of wave nature}, $N_{\rm Wave}(\hat\rho)$, of $\hat\rho$ as
the maximum of $I_{\rm Wave}^{\rm (max)}$ over all possible preparations at A.  This maximum is
given by the largest value of the bound on the right-hand side of \eqr{I_W_max}, that is, for
$\sum_j\hat\rho'_j=\hat\id$. Hence we have
\beq
  \label{N_wave}
   N_{\rm Wave}(\hat\rho)\le \log(D)-\sum_j S({\cal G}[\hat\rho])
\eeq
where $D$ is the dimension of the Hilbert space. We have previously defined the quantity
$W_G(\hat\rho)=\log(D)-S({\cal G}[\hat\rho])$ as the symmetry of $\hat\rho$ with respect to $G$
\cite{Vacc}. Thus {\it the information-theoretic measure of wave nature is bounded by the symmetry
of $\hat\rho$}, which is consistent with our association of waves with symmetry.

Combining the two expressions (\ref{N_part}) and (\ref{N_wave}) yields the complementarity
relation:
\beq
   N_{\rm Part}(\hat\rho)+N_{\rm Wave}(\hat\rho)\le \ln(D)-S(\hat\rho)\ .
\eeq
That is, {\it the sum of the information-theoretic measures of particle and wave natures is bounded
by the maximum information that can be carried by the system.}  A more extensive analysis will be
reported elsewhere \cite{elsewhere}.

\section{Acknowledgements}  I thank Prof. Howard Wiseman for
encouraging discussions.  This work was supported by the Australian Research Council and the State
of Queensland.


\vspace{-4.5mm}
\begin{thebibliography}{00}

\vspace{-2.5mm}
\bibitem{bohr}
N. Bohr, Phys. Rev. {\bf 48}, 696 (1935).

\bibitem{WooZur}
W. K. Wootters and W. H. Zurek, Phys. Rev. D {\bf 19}, 473 (1979)

\bibitem{ScuEng}
M. O. Scully, B. -G. Englert, H. Walther, Nature {\bf 351}, 111 (1991)

\bibitem{Wise}
H. M. Wiseman, Found. Phys. {\bf 28}, 1619 (1998)

\bibitem{Eng}
B. -G. Englert, Phys. Rev. Lett. {\bf 77}, 2154 (1996)

\vspace{-2.5mm}
\bibitem{Ozawa}
See e.g. M. Ozawa, Phys. Lett. A {\bf 320}, 367 (2004), and references therin.

\vspace{-2.5mm}
\bibitem{Schumacher}
B. Schumacher, M. Westmoreland, and W. K. Wootters, Phys. Rev. Lett. {\bf 76}, 3452 (1996).

\vspace{-2.5mm}
\bibitem{Vacc}
J. A. Vaccaro, F. Anselmi, H. M. Wiseman, and K. Jacobs, {\em Complementarity between extractable
mechanical work, accessible entanglement, and ability to act as a reference frame, under arbitrary
superselection rules}, quant-ph/0501121.

\vspace{-2.5mm}
\bibitem{Englert}
B. -G. Englert, Phys. Rev. Lett. {\bf 77}, 2154 (1996).

\bibitem{BarWis}
S. D. Bartlett and H. M. Wiseman, Phys. Rev. Lett. {\bf 91}, 097903 (2003)

\bibitem{elsewhere} J. A. Vaccaro
(in preparation).

\end{thebibliography}
\end{document}